\documentstyle[aps,12pt,floats,epsf,times,latexsym]{revtex}
\tighten

\def\np#1#2#3   {{ Nucl. Phys.} {\bf#1}, #2 (#3). }
\def\pcps#1#2#3 {{ Proc. Cam. Phil. Soc.} {\bf#1}, #2 (#3). }
\def\pl#1#2#3   {{ Phys. Lett.} {\bf#1}, #2 (#3). }
\def\plc#1#2#3   {{ Phys. Lett.} {\bf#1}, #2 (#3); }
\def\prep#1#2#3 {{ Phys. Rep.} {\bf#1}, #2 (#3). }
\def\prev#1#2#3 {{ Phys. Rev.} {\bf#1}, #2 (#3). }
\def\prl#1#2#3  {{ Phys. Rev. Lett.} {\bf#1}, #2 (#3). }
\def\prs#1#2#3  {{ Proc. Roy. Soc.} {\bf#1}, #2 (#3). }
\def\ptp#1#2#3  {{ Prog. Th. Phys.} {\bf#1}, #2 (#3). }
\def\rmp#1#2#3  {{ Rev. Mod. Phys.} {\bf#1}, #2 (#3). }
\def\rpp#1#2#3  {{ Rep. Prog. Phys.} {\bf#1}, #2 (#3). }
\def\zp#1#2#3   {{ Zeit. Phys.} {\bf#1}, #2 (#3). }
\def\epj#1#2#3   {{ Eur. Phys. Jour.} {\bf#1}, #2 (#3). }
\def\nim#1#2#3   {{ Nucl. Instr. Meth.} {\bf#1}, #2 (#3). }
\newcommand{\nubar}[0]{\overline{\nu}}
\newcommand{\stw}{\mbox{$\sin^2\theta_W$}}
\newcommand{\ope}{1+{1\over2}\epsilon}
\newcommand{\opbe}{1+{1\over2}\overline{\epsilon}}
\newcommand{\nub}{\overline{\nu}}

\begin{document}

%
%
%\date{\today}
%maketitle
%pacs{PACS numbers: 13.15.+g, 12.15.Mm, 12.60.Rc, 14.80.Lm}
%\ifnum\makepreprint=0
%\twocolumn
%\fi
%}

%
%*********************************************
%%%%%%%%%%%%%%%%%%% ARTICLE TEXT %%%%%%%%%%%%%%%%%%%%
%************************************

%\input{introduction.tex}
%auto-ignore
\title{Shadowing Corrections and the  
Precise Determination of Electroweak Parameters 
in Neutrino-Nucleon Scattering}
\author{Gerald A. Miller
\\ University of Washington
  Seattle, WA 98195-1560}
\author{A.~W.~Thomas\\
Department of Physics and Mathematical Physics, and Special
Research Centre for the Subatomic Structure of Matter\\
Univ. of Adelaide, Adelaide 5005, Australia}
\maketitle
\vskip0.5cm

\begin{abstract}

%%gm the 
A  systematic error in the extraction of
$\sin^2 \theta_W$ from nuclear deep inelastic scattering of neutrinos
and antineutrinos
 arises  from higher-twist effects arising from 
nuclear shadowing. We explain that these effects
cause a correction to the results of the recently reported 
significant deviation from the Standard Model 
that is  potentially as large as the
deviation claimed, and of a sign that  cannot be determined without an
extremely careful study of the data set used to model the input parton
distribution functions.
\end{abstract}
In a recent and 
stimulating paper\cite{Zeller:2001hh}, the NuTeV collaboration reported a 
determination of $\sin^2 \theta_W$, based on a comparison of
charged and neutral 
current neutrino interactions with a nuclear target (Fe),
which differs from the Standard Model prediction
by three standard deviations. In view of the importance of such a result it is
vital that the sources of systematic error be clearly identified
and examined.  Here we     
explain that there is a nuclear correction,
arising from the higher-twist effects of nuclear
shadowing\cite{Boros:1998es,Boros:1998mt},  
%%gm two refs since specific figure is cited below
 for which no allowance has been made in
the NuTeV analysis. This correction
may well be of the same size as the reported deviation.

The measurement under consideration 
involves the
separate  measurements of the ratios of 
neutral current (NC) to charged current (CC) cross sections on Fe
{}for $\nu$ and $\bar \nu$. The best values of $\stw$ and $\rho_0$
are extracted from the precisely determined ratios. But the nuclear effects
must be removed. Because 
a substantial fraction of 
the NuTeV data in the region $x$ below 0.1 is at relatively low
$Q^2$ (even though the average $Q^2$ is 16 GeV$^2$), one expects
a significant shadowing contribution from vector meson 
dominance (VMD)\cite{Kwiecinski:ys,Melnitchouk:1995am},
which is of higher twist. As
explained  by Boros {\it et al.}\cite{Boros:1998es,Boros:1998mt}, 
the effect of the VMD contribution to nuclear
shadowing in neutrino interactions
is substantial, and 
leads to a reduction of the CC
$\nu$ cross section by
about 50\% compared with the reduction found for photons.
(Briefly, the VMD 
contribution to shadowing is
dominated by the $\rho$ meson and $f^2_{\rho^+} = 2
f^2_{\rho^0}$,
whereas the CC to photon cross sections are in the ratio 
18/5.) Together with 
a full NLO analysis of the data, 
this was important in reconciling the NuTeV and 
NMC data without any need for
substantial charge symmetry violation of the parton 
distributions \cite{Boros:1999fy}. A recent re-examination of 
the role of vector meson dominance in nuclear shadowing at low $Q^2$ finds
that models (such as that used here)
which incorporate both vector meson and partonic mechanisms are
consistent with both the magnitude and the $Q^2$ slope of the shadowing  
data~\cite{Melnitchouk:2002ud}.

{}For present purposes we need also to consider
this higher-twist effect of shadowing of
vector mesons for  $\nubar$ interactions. These involve predominantly 
anti-quarks and the shadowing effect is relatively
larger by a factor of three or so.
However, the VMD contribution to shadowing 
for neutral current interactions is 1/2 of that for charged current 
interactions because 
$Z$ conversion to a $\rho^0$ occurs with a
{}factor of $(1/2)$ of that for $W^+\to\rho^+$.

Let us examine how these differences in shadowing effects influence the
extraction of $\stw$. Suppose the
nuclear cross section for NC interactions of neutrinos is
larger than that for CC interactions by a factor of $\ope$
and that the one for anti-neutrinos is larger by a factor of
$\opbe$, with $\overline{\epsilon}$ expected to be
substantially larger than
$\epsilon$.
Then the nuclear ratios of neutral current (NC)
to charged current (CC) cross sections are 
\begin{eqnarray}
&&  R^{\nu}_A \equiv \frac{\sigma_A(\nu A\rightarrow\nu X)}
                 {\sigma_A(\nu A\rightarrow\ell^{-}X)}  
= \frac{\sigma(\nu N\rightarrow\nu X)}
                 {\sigma(\nu N\rightarrow\ell^{-}X)}(\ope)
                 = (\ope)(g_L^2+r\;g_R^2),\\
&& R^{(\nub)}_A \equiv \frac{\sigma_A(\nub A\rightarrow\nub X)}
                 {\sigma_A(\nub A\rightarrow\ell^{(+)}X)}  
= (\opbe)(g_L^2+r^{(-1)}g_R^2),
\label{eqn:ls}
\end{eqnarray}
where
%\begin{equation}
$r= {\sigma({\overline \nu}N\rightarrow\ell^+X)}/
                {\sigma(\nu N\rightarrow\ell^-X)}
                ,\;r_A=r(\ope)/(\opbe),\;
$
%\end{equation}
%%The neutrino-quark coupling constants are given by
$g_L^2=1/2-\stw+5/9\sin^4\theta_W$ and $g_R^2= 5/9\sin^4\theta_W$.
Equations (1) and (2) tell us that
the nuclear-shadowing corrections for $ R^{\nu}_A$ and $R^{(\nub)}_A$ are not
the same, and that  
the extraction of $\stw$ requires the
separate knowledge  of  
$\epsilon$ and $\overline{\epsilon}$.

A detailed analysis of the NuTeV data requires that one model the ratios
$ R^{\nu}_A$ and $R^{(\nub)}_A$ at a %%gm
required accuracy of a  fraction of a percent. This, in turn,
requires an %%gm
even more  accurate knowledge of both the quark and antiquark
parton distribution functions (pdfs). In general, the pdfs are derived
from a global analysis data from electron/muon, CC neutrino and NC
neutrino deep inelastic scattering on protons, deuterons and nuclei. The
range of $Q^2$, particularly at low $x$ ($x \leq 0.1$) can be quite
low. Higher twist shadowing corrections are almost universally ignored
in global determinations of the pdfs. This is certainly the case for the
pdfs used by NuTeV. Since the VMD shadowing corrections are different
for electrons, CC neutrino scattering and NC neutrino scattering, the
pdfs resulting from such a global analysis are at best an approximation
to the true ones, with unknown systematic errors. 
Worse, one cannot simply add a shadowing correction to
a simulation based on such global pdfs as even the sign of the
correction will depend on the particular data sets included in the
analysis. Of course, the systematic errors
encountered will not be serious for most purposes. 
However, in this case, where the
signal of a deviation from the Standard Model is at the percent level,
one must control this potential source of error extremely carefully.

We cannot undo the global analysis of pdfs by the NuTeV collaboration.
However, we can make an estimate of the order of magnitude of the
shadowing corrections using Eqs.~(1) and (2).
The quantity $\epsilon$ is given as a 
{}function of $x$ for $Q^2=5 \;{\rm GeV}^2$
as the dashed curve of  Fig~ 3b. in Ref.~\cite{Boros:1998es}.
One needs to take the deviation between dashed curve and unity.
Thus, e.g. 
$\epsilon(x,Q^2= 5{\rm GeV}^2)\approx 0.041$ at $x=10^{-2}$. For other
values of $Q^2$, one may use:
$\epsilon (x,Q^2)\approx \epsilon(x,Q^2= 
5{\rm GeV}^2) \left( {m_\rho^2+5 {\rm GeV}^2\over
m_\rho^2+Q^2} \right)^2 $. Furthermore
$\overline\epsilon/{\epsilon}\approx F_2^D(\nu)/ F_2^D(\nub)\approx2.$ 
Using a typical value  of $Q^2\approx10 GeV^2$  and range of $x\approx 0.05$
of the NuTeV  %%gm
experiment leads to
the values $\epsilon=0.006$ and $\overline{\epsilon}=0.012$\cite{kevin}.
Thus the nuclear corrected ratios $ R^{\nu}_A$ and $ R^{\nub}_A$ would be
smaller than those reported in Ref.~\cite{Zeller:2001hh}, by 0.003 and 0.006
respectively. That these numbers represent important
corrections can be seen immediately  by examining the sources of errors
reported in Ref.~\cite{Zeller:2001hh}.
The total error for $ R^{\nu}_A$  is reported as 0.0013 and that for 
$ R^{\nub}_A$ as 0.00272. The effects of shadowing are larger 
than these quoted errors by a
{}factor of two or three!   It is clear that any analysis of nuclear data aimed
at determining $\stw$ must account for nuclear shadowing. But this has not
been done in  Ref.~\cite{Zeller:2001hh}.

It is  necessary to carry out
a more refined analysis of the data which properly
incorporates the high twist components of nuclear shadowing, starting
with the pdfs themselves. Such an analysis should also take
into account the experimental acceptances as a
function of $x$, $y$ and $Q^2$.
Alternatively, one could drastically reduce the VMD contribution
by restricting the data set to events with $Q^2>5\; {\rm GeV}^2$ --
although we understand that this may present difficulties for NC events.

We find  that the size of the shadowing effects is substantial and
should be incorporated in the experimental analysis. It is true that
the simplest estimate of this  effect (ignoring the effects on the pdfs
themselves, which was discussed above) 
is opposite to that required to explain the deviation from the
Standard Model. Thus it may be that the deviation 
{}from the Standard Model could be even {\em larger} than reported in
Ref.~\cite{Zeller:2001hh}.

{}Finally, we note that
several other effects that tend to reduce the discrepancy have been 
reported. The influence of charge symmetry breaking, arising from the mass
difference between up and down quarks\cite{csb}, accounts for about a third to
a half of the deviation between NuTeV's value of $\stw$ and that of the
Standard Model in a model-independent manner\cite{ncsb} .
{}Furthermore, it has been known for more 
than 20 years that parton distributions
of nucleons bound in nuclear matter differ from those of free nucleons.
Such effects\cite{emc} still present a considerable challenge to our
understanding of nonperturbative QCD and it is not inconceivable that
they could eventually account for the entire deviation of  $\stw$.

It seems clear that the 
extraction of the the value of $\stw$ from neutrino-nuclear interactions 
involves handling several different types of corrections of different signs,
including some
that are difficult evaluate with precision.
The situation here may well be similar
to many in strong interaction physics, in which a ``cocktail''
of effects is required\cite{taubes}. 
Considering that possible explanations in terms of 
new physics are not compelling\cite{Davidson:2002fb}, considerable efforts
must be applied before concluding that the NuTeV result really 
demonstrates a deficiency of the Standard Model.

We thank the USDOE for partial support and M.~Ramsey-Musolf for a useful remark.
This work was also supported by the Australian Research Council.


\begin{references}
%\cite{Zeller:2001hh}
\bibitem{Zeller:2001hh}
G.~P.~Zeller {\it et al.}, %%gm  [NuTeV Collaboration],
%``A precise determination of electroweak 
%parameters in neutrino nucleon  scattering,''
Phys.\ Rev.\ Lett.\  {\bf 88}, 091802 (2002).
%%gm[arXiv:hep-ex/0110059].
%%CITATION = HEP-EX 0110059;%%
%
\bibitem{Boros:1998es}
C.~Boros, J.~T.~Londergan and A.~W.~Thomas,
%``Evidence for charge symmetry violation in parton distributions,''
Phys.\ Rev.\ D {\bf 59}, 074021 (1999);
%%gm [arXiv:hep-ph/9810220];
\bibitem{Boros:1998mt}
%%C.~Boros, J.~T.~Londergan and A.~W.~Thomas,
%``Shadowing in neutrino deep inelastic scattering and the determination
%of the strange quark distribution,''
Phys.\ Rev.\ D {\bf 58}, 114030 (1998).
%%gm[arXiv:hep-ph/9804410].
%%CITATION = HEP-PH 9804410;%%
%
\bibitem{Kwiecinski:ys}
J.~Kwiecinski and B.~Badelek,
%``Unified Description Of Nuclear Shadowing Of Virtual Photons,''
Phys.\ Lett.\ B {\bf 208}, 508 (1988).
%%CITATION = PHLTA,B208,508;%%
%\cite{Melnitchouk:1995am}
\bibitem{Melnitchouk:1995am}
W.~Melnitchouk and A.~W.~Thomas,
%``Q**2 dependence of nuclear shadowing,''
Phys.\ Rev.\ C {\bf 52}, 3373 (1995).
%%gm[arXiv:hep-ph/9508311].
%%CITATION = HEP-PH 9508311;%%
%
\bibitem{Boros:1999fy}
C.~Boros, F.~M.~Steffens, J.~T.~Londergan and A.~W.~Thomas,
%``A new analysis of charge symmetry violation in parton distributions,''
Phys.\ Lett.\ B {\bf 468}, 161 (1999).
%%gm[arXiv:hep-ph/9908280].
%%CITATION = HEP-PH 9908280;%%

%\cite{Melnitchouk:2002ud}
\bibitem{Melnitchouk:2002ud}
W.~Melnitchouk and A.~W.~Thomas,
%``Nuclear shadowing at low Q**2 and the extraction of sin**2 Theta(W),''
Phys.\ Rev.\ C {\bf 67}, 038201 (2003)
[arXiv:hep-ex/0208016].
%%CITATION = HEP-EX 0208016;%%
\bibitem{kevin} %%gmK.~S. McFarland, and G.~P. Zeller, private communication.
For accuracy, one needs to know the  fraction of the NuTeV events obtained 
 at low $x$ and $Q^2$. Using $Q^2=10$ GeV$^2$ reflects a compromise
between using  using the average value $\sim20$ GeV$^2$ where the shadowing
is completely absent, and  the lower values  characteristic of the low $x$
region where 
 shadowing is important. 


\bibitem{csb} G.~A.~Miller, B.~M.~Nefkens and I.~Slaus,
%``Charge Symmetry, Quarks And Mesons,''
Phys.\ Rept.\  {\bf 194}, 1 (1990).
%%CITATION = PRPLC,194,1;%%
\bibitem{ncsb}E.~Sather,
%``Isospin violating quark distributions in the nucleon,''
Phys.\ Lett.\ B {\bf 274}, 433 (1992);
%%CITATION = PHLTA,B274,433;%%
E.~N.~Rodionov, A.~W.~Thomas and J.~T.~Londergan,
%``Charge Asymmetry Of Parton Distributions,''
Mod.\ Phys.\ Lett.\ A {\bf 9}, 1799 (1994);
%%CITATION = MPLAE,A9,1799;%%
J.~T.~Londergan and A.~W.~Thomas,
%``Charge symmetry violation corrections to determination of the Weinberg  angle in neutrino reactions,''
Phys.\ Rev.\ D {\bf 67}, 111901 (2003)
[arXiv:hep-ph/0303155];
%%CITATION = HEP-PH 0303155;%%
J.~T.~Londergan and A.~W.~Thomas,
%``Charge symmetry violating contributions to neutrino reactions,''
Phys.\ Lett.\ B {\bf 558}, 132 (2003)
[arXiv:hep-ph/0301147].
%%CITATION = HEP-PH 0301147;%%
\bibitem{emc}
S.~Kovalenko, I.~Schmidt and J.~J.~Yang,
%``Nuclear effects on the extraction of sin**2(Theta(W)),''
Phys.\ Lett.\ B {\bf 546}, 68 (2002)
[arXiv:hep-ph/0207158];
%%CITATION = HEP-PH 0207158;% 
S.~Kumano,
%``Modified Paschos-Wolfenstein relation and extraction of weak mixing  angle sin**2(theta(W)),''
Phys.\ Rev.\ D {\bf 66}, 111301 (2002)
[arXiv:hep-ph/0209200].
%%CITATION = HEP-PH 0209200;%%
S.~A.~Kulagin,
%``Paschos-Wolfenstein relationship for nuclei and the NuTeV  sin**2(theta(W)) measurement,''
Phys.\ Rev.\ D {\bf 67}, 091301 (2003)
[arXiv:hep-ph/0301045].
%%CITATION = HEP-PH 0301045;%%
%%gm
\bibitem{taubes} G. Taubes, ``Nobel Dreams:Power Deceit 
and the Ultimate Experiment'', Random House NY, NY (1993) 
%\cite{Davidson:2002fb}
%\cite{Davidson:2002fb}
\bibitem{Davidson:2002fb}
S.~Davidson,
%``Interpretations of the NuTeV sin**2(theta(W)),''
J.\ Phys.\ G {\bf 29}, 2001 (2003)
[arXiv:hep-ph/0209316].
%%CITATION = HEP-PH 0209316;%%
\end{references}
\end{document}